\documentclass{webofc}

\usepackage[varg]{txfonts}   
\usepackage{hyperref}
\usepackage{url}
\hypersetup{colorlinks=true,citecolor=blue,urlcolor=blue,linkcolor=blue}

\newcommand{\beq}{\begin{equation}}
\newcommand{\eeq}{\end{equation}}
\newcommand{\bea}{\begin{eqnarray}}
\newcommand{\eea}{\end{eqnarray}}

\begin{document}
\title{Nuclear Fragmentation at the Future Electron-Ion Collider}
%
%

\author{\firstname{Carlos A.} \lastname{Bertulani}\inst{1,2,3}\fnsep\thanks{\email{carlos.bertulani@etamu.edu}} 
}

\institute{Department of Physics and Astromomy, East Texas A\&M University, 
Commerce, TX 75429, USA 
\and
Institut f\"ur Kernphysik,  Technische Universit\"at Darmstadt, 64289 
Darmstadt, Germany 
\and
Helmholtz Research Academy Hesse for FAIR, D–64289 Darmstadt, Germany
          }

\abstract{We investigate aspects of low-energy nuclear reactions that could be explored at the forthcoming Electron–Ion Collider (EIC) at Brookhaven National Laboratory and compare them with analogous measurements performed in ultraperipheral collisions (UPCs) at the Large Hadron Collider (LHC) at CERN.
The estimated fragmentation cross sections at the EIC are roughly three orders of magnitude smaller than those observed at the LHC.
At the LHC, uranium nucleus fragmentation exhibits a distinctive double-peaked mass spectrum arising from fission processes, whereas at the EIC, the breakup pattern is mainly characterized by neutron evaporation and a vastly reduced yield of fission fragments, about four orders of magnitude fewer events in comparison.
}
\maketitle
\section{Introduction}

Fragments generated through Coulomb excitation of fast-moving nuclei in ultra-peripheral collisions (UPCs) have yielded valuable insights into fission mechanisms, the identification of novel isotopes and isomeric configurations, and the characterization of over a thousand fission residues \cite{SCHMIDT2002157,JURADO2003186,JURADO200514}.
This technique represents a powerful probe for exploring the fission pathway, neutron evaporation, and their connection to the astrophysical rapid neutron capture (r-)process, as well as other modes of nuclear decay \cite{PEREZLOUREIRO2011552}.
The fragmentation mechanism is primarily triggered by the excitation of collective modes, particularly the giant dipole resonance (GDR), which sensitively reflects the isospin asymmetry of the nucleus.
Such observables can provide constraints on the symmetry energy slope parameter $L$ within the nuclear matter equation of state (EoS) \cite{ROCAMAZA201896,TamiPRL.107.062502}.

Beyond the UPC environment in relativistic heavy-ion collisions, it is of great interest to assess how the Electron–Ion Collider (EIC) might be exploited to probe excitation and decay dynamics, and whether it could enable the synthesis of nuclei unreachable in heavy-ion reactions.
This serves as the principal motivation for the work published in Ref.  \cite{BYN2025}, which addresses the fragmentation of nuclei induced by electromagnetic (EM) interactions with electrons at sub-50 MeV photon energies ($\hbar \omega < 50$ MeV).
The dominant energy window for photon-induced nuclear fragmentation corresponds to the GDR region, typically around 10–20 MeV for heavy systems and somewhat higher for lighter ones.
Electron–nucleus scattering exhibits substantial cross sections for the excitation of GDRs or other high-lying collective modes, which subsequently decay via the emission of neutrons, protons, light charged particles, or, in the case of actinides, fission fragments.
Neutrons produced in this mechanism are low in energy in the ion rest frame, far below those emerging from central relativistic heavy-ion collisions, and hence serve as sensitive markers of intense EM fields.
In the laboratory frame, these neutrons attain energies comparable to the ion’s kinetic energy per nucleon.
We show   \cite{BYN2025} that neutron emission and fission processes, primarily arising from EM dissociation via GDR decay, are of considerable importance and could be extensively detected at the future EIC.
We further contrast the fragment yields predicted for the EIC with those expected from heavy-ion UPCs at the Large Hadron Collider (LHC) at CERN \cite{BERTULANI1988299}.
Our findings emphasize the role of various multipolarities, the relation between the virtual photon spectra at the EIC and in UPCs, and the distinct features that characterize low-energy photonuclear phenomena in these two experimental regimes  \cite{BYN2025}.

\section{EM Fragmentation at the LHC and at the EIC}
To evaluate the excitation of the isovector giant dipole resonance (IVGDR) and the isoscalar and isovector giant quadrupole resonances (ISGQR and IVGQR)
\cite{BERTULANI1988299,aumann1995ZP,BERTULANI1999139,PhysRevLett.124.132301},
we adopt a Lorentzian distribution to represent each mode.
For the IVGDR, the centroid energy is taken as
$\omega_{0} = 31.2A^{-1/3} + 20.6A^{-1/6}$ MeV, and its strength parameter $\sigma_0$ is fixed through the Thomas–Reiche–Kuhn (TRK) sum rule, a nearly model-independent expression for the electric dipole nuclear response \cite{eisenberg1988excitation}.
The ISGQR and IVGQR centroids are defined by $\omega_0 = 62A^{-1/3}$ MeV and $130A^{-1/3}$ MeV, respectively, with all resonances assumed to saturate their associated operator sum rules, as reported in Ref. \cite{Harakeh:02}.
We consider an 18 GeV electron beam interacting with a 110 GeV/nucleon
${}^{238}\text{U}$ or ${}^{208}\text{Pb}$ beam at the EIC, corresponding to a laboratory energy of $E_{\text{lab}} = 89$ GeV.
For comparison, we examine ${}^{208}\text{Pb} + {}^{208}\text{Pb}$ collisions at the LHC, where $\sqrt{s_{NN}} = 5.5$ TeV/nucleon, yielding $E_{\text{lab}} = 2.76$ TeV/nucleon and a Lorentz factor $\gamma_{\text{lab}} = 2941$.
A hypothetical case of ${}^{238}\text{U} + {}^{238}\text{U}$ at the LHC is also analyzed to explore fission-channel fragment production under both accelerator conditions.

De-excitation of the giant resonances results in nucleon and light-charged particle emission, gamma decay, intermediate-mass fragments (IMFs), and fission residues.
The cross section for producing an isotope $x = (Z, A)$ is expressed as
$
\sigma_x(\omega) = b_x(\omega) \sum_{GR} \sigma_\gamma^{GR}(\omega),
$
where the sum runs over all giant resonances (GRs) and $b_x(\omega) = 
{\Gamma_x(\omega^*)}/{\Gamma_{tot}(\omega^*)}$ denotes the branching ratio for emission of fragment $x$ at excitation energy $\omega$. Here, $\Gamma_x$ represent the partial and total decay widths, respectively, and for giant resonance excitation, pre-equilibrium emission is negligible so that $\omega^* \approx \omega$.
The branching ratio $b_x(\omega)$ is evaluated using the Ewing–Weisskopf model.
Separation energies and Coulomb barriers for charged particles are included following the 2017 atomic mass evaluation \cite{ame2016a} and Bass potential \cite{bass} for transmission probability calculations.
Fission yields are obtained via the dynamical approach detailed in
Refs. \cite{JURADO2003186,JURADO200514}.

The mass spectrum of Pb fragments at the LHC shows a rapid falloff with decreasing fragment mass, primarily due to neutron evaporation.
A similar pattern appears for uranium, although fission fragments originating from ${}^{238}\text{U}$ excitation yield a characteristic double-humped structure with maxima around $A \approx 100$ and $A \approx 140$, indicative of fission splitting.
At the EIC, the corresponding cross sections are roughly $10^3$ times smaller.
Both nuclei exhibit steeply declining fragment distributions toward lighter masses due to neutron loss, and the fission peaks near $A \sim 100$ and $A \sim 140$ persist for uranium.
Neutron removal proceeds much faster at the EIC than in UPCs at the LHC, with nearly all events dominated by few-neutron evaporation and occasional fission residues  \cite{BYN2025}.

Table~\ref{tab1} summarizes the calculated cross sections for
1–4 neutron evaporation channels ($\sigma_{-1n}$, $\sigma_{-2n}$, $\sigma_{-3n}$, $\sigma_{-4n}$), fission, total yields, and fission branching ratios.
The LHC values (in barns) exceed the EIC results (in millibarns) by about three orders of magnitude, consistent with the expected scaling of electromagnetic excitation probabilities.
In both settings, neutron emission dominates, while light-particle channels contribute only marginally and the mass distribution drops sharply.
For ${}^{238}$U, fission residues appear at both facilities.
Because electromagnetic excitation energies are modest (around 15 MeV), much smaller than those induced by strong interactions \cite{PhysRevLett.124.132301}, the cross sections for neutron-rich isotope production remain comparatively low.
In contrast, hard lepton–nucleus interactions, involving initial lepton–parton scattering followed by an intra-nuclear cascade \cite{magdy2024}, tend to generate neutron-deficient isotopes in the $Z = 89$–94 region \cite{Schmookler2023}.
Nevertheless, their total cross sections are below those obtained for collective excitation of giant modes, since the lepton–parton interaction strength is inherently weaker than the coherent electromagnetic nuclear response.

Overall, our analysis indicates that UPCs at the LHC are likely more favorable for nuclear-structure investigations, provided appropriate experimental detection conditions are achieved.
The EIC, however, offers the benefit of a clean electron probe, potentially allowing clearer identification of nuclear fragments and complementary information about electromagnetic dissociation dynamics  \cite{BYN2025}.

  \begin{table}[ht]
  \centering
  \caption{Cross sections, expressed in barns for LHC  and in millibarns for EIC, are presented for various decay pathways of lead and uranium nuclei studied at both the LHC and the forthcoming EIC.
The table includes results for neutron evaporation channels ($-1n$, $-2n$, $-3n$, and $-4n$), together with fission probabilities, the total integrated cross sections, and the fractional contributions of the fission branch to the overall reaction yield. }
    \label{tab1}
 \begin{tabular}{|c|c|c|c|c|}
 \hline
 Cross sections     & LHC & LHC & EIC &EIC  \\
 \hline
  & Pb + Pb [b]& U + U [b] & e-Pb [mb] &e-U [mb]  \\
    \hline
    \hline
    $\sigma_{-1n}$ & 33.93 & 33.20&  20.24& 15.58\\
    $\sigma_{-2n}$ & 18.89 &30.59&  11.45& 14.88\\
    $\sigma_{-3n}$ & 2.546 &3.537&  1.416& 1.591\\
     $\sigma_{-4n}$ & 1.091 &0.784&  0.5933& 0.2934\\
     $\sigma_{fission}$ & 0 &18.24&  0& 8.867\\
     \hline
     $\sigma_{total}$ & 55.74 &85.48&  33.90& 41.32 \\  
     \hline  
Fission b.r.& 0\% &19.54\%&  0\%& 21.45\%\\   
\hline
    \end{tabular}
  \end{table}

At the EIC, nuclear excitation and decay can occur through initial photon–nucleon interactions followed by secondary partonic or nucleonic cascades.
However, the associated cross sections remain comparatively small relative to those generated via the collective excitation of giant resonances  \cite{BYN2025}.
In contrast, at heavy-ion colliders such as the LHC, nuclear fragmentation may also originate from central collisions, where the system can evolve into a hot, dense fireball, possibly undergoing a QCD phase transition before hadronization.
Our comparative analysis of electromagnetic fragmentation at the EIC and LHC thus represents two distinct limiting scenarios, while also emphasizing the relevance of future high-intensity laser facilities in probing analogous excitation and dissociation mechanisms \cite{negoita2022,PhysRevAccelBeams.25.101601}.

Unlike $eA$ collisions at the EIC, ultraperipheral heavy-ion collisions (UPCs) may involve pomeron exchange, a long-range strong-interaction process that competes with photon-mediated reactions and can likewise contribute to nuclear breakup.
Nevertheless, these mechanisms are fundamentally distinct: pomerons carry quantum numbers $0^{++}$, whereas the photon possesses $1^{--}$.
Because pomeron contributions are absent in the EIC environment, it provides a unique setting to investigate purely electromagnetic excitation processes, isolated from strong-interaction effects, in studies of low-excitation nuclear dynamics.

\medskip

{\it Acknowledgement.}  This work has been supported by the U.S. DOE grant DE-SC0026074 and  the ExtreMe Matter Institute EMMI at the GSI Helmholtzzentrum für Schwerionenforschung, Darmstadt, Germany.

\end{document}